\documentclass{elsart}

\usepackage[square,comma]{natbib}
\usepackage{graphicx}
\usepackage{amssymb}

\begin{document}

\thispagestyle{empty}
\begin{Large}
\textbf{DEUTSCHES ELEKTRONEN-SYNCHROTRON}

\textbf{\large{in der HELMHOLTZ-GEMEINSCHAFT}\\}
\end{Large}

DESY 15-021

February 2015

\begin{eqnarray}
\nonumber
\end{eqnarray}
\begin{center}
\begin{Large}
\textbf{Theoretical computation of the polarization characteristics of an X-ray Free-Electron Laser with planar undulator}
\end{Large}
\begin{eqnarray}
\nonumber &&\cr \nonumber && \cr
\end{eqnarray}

\begin{large}
Gianluca Geloni,
\end{large}
\textsl{\\European XFEL GmbH, Hamburg}
\begin{large}

Vitali Kocharyan and Evgeni Saldin
\end{large}
\textsl{\\Deutsches Elektronen-Synchrotron DESY, Hamburg}

\begin{eqnarray}
\nonumber
\end{eqnarray}
\begin{eqnarray}
\nonumber
\end{eqnarray}
ISSN 0418-9833
\begin{eqnarray}
\nonumber
\end{eqnarray}
\begin{large}
\textbf{NOTKESTRASSE 85 - 22607 HAMBURG}
\end{large}
\end{center}
%\end{widetext}
\clearpage
\newpage

\begin{frontmatter}



\title{Theoretical computation of the polarization characteristics of an X-ray Free-Electron Laser with planar undulator}


\author[XFEL]{Gianluca Geloni}
\author[DESY]{Vitali Kocharyan}
\author[DESY]{Evgeni Saldin}

\address[XFEL]{European XFEL GmbH, Hamburg, Germany}
\address[DESY]{Deutsches Elektronen-Synchrotron (DESY), Hamburg,
Germany}

\begin{abstract}

We show that radiation pulses from an X-ray Free-Electron Laser (XFEL) with a planar undulator, which are mainly polarized in the horizontal direction, exhibit a suppression of the  vertical polarization component of the power at least by a factor $\lambda_w^2/(4 \pi L_g)^2$, where $\lambda_w$ is the length of the undulator period and $L_g$ is the FEL field gain length. We illustrate this fact by examining the XFEL operation under the steady state assumption. In our calculations we considered only resonance terms: in fact, non resonance terms are suppressed by a factor $\lambda_w^3/(4 \pi L_g)^3$ and can be neglected. While finding a situation for making quantitative comparison between analytical and experimental results may not be straightforward, the qualitative aspects of the suppression of the vertical polarization rate at XFELs should be easy to observe. We remark that our exact results can potentially be useful to developers of new generation FEL codes for cross-checking their results.

\end{abstract}


\end{frontmatter}

% main text

\clearpage
\section{\label{sec:intro} Introduction}

A Free-Electron Laser (FEL) amplifier which starts up from the shot noise in the electron beam is known as a self-amplified spontaneous emission (SASE) FEL. In the SASE case, the amplification process has its origins in the density fluctuations of the electron beam. SASE FELs are capable of producing coherent, tunable FEL radiation down  to a fraction of an Angstrom \cite{LCLSI,SACLAI}. A SASE FEL which operates in the X-ray wavelength range is called XFEL.

With the advent of FELs,  X-ray radiation pulses with unprecedented characteristics were made available to the scientific community. Compared to conventional synchrotron radiation sources, X-ray FELs (XFELs) offer an increase in peak brightness of many orders of magnitude as well as ultrashort pulses in the femtosecond time scale. In this paper we consider an additional feature of XFEL pulses, which is useful in many experiments. Namely the fact that the  pulses produced by an XFEL with horizontal planar undulator exhibit an  extremely small component of the electric field in the vertical direction. In particular, here we will show that for a typical XFEL setup the horizontally polarized component of radiation is greatly dominant, and that only less that one part in a million of the total intensity is polarized in the vertical plane.

The study of XFEL polarization characteristics is obviously deeply related to the problem of electromagnetic wave amplification in XFEL, which refers to a particular class of self-consistent problems. It can be separated into two parts: the solution of the dynamical problem, i.e. finding the motion of the electrons in the beam under the action of given electromagnetic fields, and the solution of the electrodynamic problem, i.e. finding the electromagnetic fields generated by a given contribution of charge and currents. The problem is closed by simultaneous solution of the field equations and of the equations of motion.

Let us consider the electrodynamic problem more in detail. The equation for the electric field follows from Maxwell's equations. One obtains, in Gaussian units:

\begin{equation}
c^2 \vec{\nabla}\times(\vec{\nabla}\times{\vec{E}}) = -
\frac{\partial^2 \vec{E}}{\partial t^2} - 4 \pi \frac{\partial
\vec{j}}{\partial t}~. \label{elec}
\end{equation}
With the help of the identity

\begin{equation}
\vec{\nabla}\times(\vec{\nabla}\times{\vec{E}}) =
\vec{\nabla}(\vec{\nabla}\cdot{\vec{E}})-\nabla^2
\vec{E}\label{iden}
\end{equation}
and Poisson equation
\begin{equation}
\vec{\nabla}\cdot\vec{E} = 4 \pi \rho \label{pois}
\end{equation}
we obtain the inhomogeneous wave equation for $\vec{E}$

\begin{equation}
c^2 \nabla^2 \vec{E} - \frac{\partial^2 \vec{E}}{\partial t^2} = 4
\pi c^2 \vec{\nabla} \rho + 4 \pi \frac{\partial \vec{j}}{\partial
t}~. \label{diso}
\end{equation}
Once the charge and current densities  $\rho$ and $\vec{j}$ are specified as a function of time and position, this equation allows one to calculate the electric field $\vec{E}$ at each point of space and time \cite{jackson99}. The current density source provides the main contribution to the radiation field in an FEL amplifier, and the contribution of the charge density source to the amplification process is negligibly small. This fact is commonly known and accepted in the FEL community. However, we have been unable to find a proof of this fact in literature, except book \cite{FELB} and review \cite{HUAKIM}, which are only the  publications, to the authors' knowledge, dealing with this issue.

Due to linearity, without the gradient term the solution of Eq. (\ref{diso}) exhibits the property that the radiation field  $\vec{E}$ points in the same direction of the current density $\vec{j}$. An important limitation of such approximation arises when we need to quantify the linear vertical field generated in the case of an XFEL with planar undulator.  In the case  $\vec{j}$ points in the horizontal direction (for a horizontal planar undulator), according to  Eq. (\ref{diso}), which is exact, only the  charge term is responsible for a vertically polarized component of the field: if it is neglected, one cannot quantify the linear vertical field anymore.

Similar to the process of harmonic generation, the process of generation of the vertically polarized field component can be considered as a purely electrodynamic one. In fact, the vertically polarized field component is driven by the charge source, but the bunching contribution due to the interaction of the electron beam with the radiation generated by such source can be neglected. This leads to important simplifications. In fact, in order to perform calculations of the radiation including the vertically  polarization component one can proceed first by solving the self-consistent problem with the current source only. This can either be done in an approximated way using an analytical model for the FEL process or, more thoroughly, exploiting  any existing FEL code. Subsequently, the solution to the self-consistent problem can be used  to calculate the first harmonic contents of the electron beam density distribution. These contents enter as  known sources in our electrodynamic process, that is Eq. (\ref{diso}). Solution of that equation accounting for these sources gives the desired polarization characteristics.

Analytical descriptions allow for a proper understanding of the physical principles underlying the phenomena under study, and also provide convenient testing of numerical simulation codes. A SASE XFEL is rather difficult to be described in a fully analytical way. In all generality,  its radiation can be represented as a non-stationary random process, and its analytical description is further complicated by the fact that the electron bunch combines both the features of the input signal and of the active medium with time-dependent parameters.

Approximations particularly advantageous for our theoretical analysis include the modeling of the electron beam density as uniform, and the introduction of a monochromatic seed signal. Realistic conditions satisfying these assumptions are the use of a sufficiently long electron bunch with a longitudinal stepped profile and the application of a scheme in the SASE mode of operation for narrowing down the radiation bandwidth. In the framework of this model it becomes possible to describe analytically all the polarization properties of the radiation from an XFEL \footnote{Our model can be close to real situations, based on two techniques recently realized at the LCLS. The first is a technique for producing a uniform electron bunch, and is heavily relying on the use of a slotted spoiler foil in the last bunch compressor chicane \cite{DING}. The method takes advantages of the high sensitivity of the FEL gain process to the transverse emittance of the electron bunch. By spoiling the emittance of most of the long nonuniform bunch while leaving short unspoiled temporal slice, one can produce electron bunch with relatively uniform active medium. The second technique is self-seeding, an active filtering technique allowing to narrow the FEL bandwidth down to almost the Fourier limit \cite{EMNAT}. Combination of these two techniques is now widely used at the LCLS.}.

The simplicity of our model offers the opportunity for an almost completely analytical description of an XFEL in the linear mode. As remarked above, a complete description of the operation of an XFEL can be performed only with time-dependent numerical simulation codes. Application of the numerical calculations allows one to describe the most general situation, including arbitrary electron beam quality and nonlinear effects. Finding an analytical solution is always fruitful for testing numerical simulation codes. Up to now, in conventional FEL codes, the contribution of the the charge source assumed to be negligible small. However, the charge term alone is responsible for the vertically polarized radiation component, which is our subject of interest. Our analytical results for the high-gain linear regime are expected to serve as a primary standard for testing future FEL codes upgrades.

\section{\label{sec:ours} XFEL radiation in resonance approximation}

As has been discussed in the Introduction, a quantification of the linear vertical field expected in FELs can be made on the basis of an electrodynamical model. We will restrict our attention to FELs driven by planar undulators. A more extended treatment of higher harmonic numbers is conceptually identical to the treatment in this paper, and only differs as regards actual calculations. Proper initial conditions following e.g. from start-to-end simulations are given as input to an FEL self-consistent code, which calculates the electron beam bunching from the interaction of the beam with the first harmonic radiation. As discussed above, in FEL codes the first harmonic is conventionally calculated accounting only for the leading terms under the resonant approximation, which yields perfectly linearly polarized radiation. Therefore, the output of these codes cannot be directly used to quantify the linear vertical field of FELs. However, the electron beam bunching can be inserted into the field equation as sources. A relatively simple electrodynamical model can then be developed in order to calculate corrections to the leading resonant terms, and therefore to calculate the linear vertical field.

Paraxial Maxwell's equations in the space-frequency domain can be used to describe radiation from ultra-relativistic electrons (see e.g. \cite{OURF}). Let us define temporal Fourier transform pairs as:

\begin{eqnarray}
&&\bar{f}(\omega) = \int_{-\infty}^{\infty} dt~ f(t) \exp(i \omega t ) \leftrightarrow
f(t) = \frac{1}{2\pi}\int_{-\infty}^{\infty} d\omega \bar{f}(\omega) \exp(-i \omega t) ~
\label{ftdef2}
\end{eqnarray}
and let us call the transverse electric field in the space-frequency domain, i.e. the Fourier transform of the real electric field in the time domain, $\vec{\bar{E}}_\bot(z,\vec{r}_\bot,\omega)$, where $\vec{r}_\bot = x\vec{e}_x+y\vec{e}_y$ identifies a point on a transverse plane at longitudinal position $z$, $\vec{e}_x$ and $\vec{e}_y$ being unit vectors in the transverse $x$ and $y$ directions. Here the frequency $\omega$ is related to the wavelength $\lambda$ by $\omega = 2 \pi c/\lambda$, $c$ being the speed of light in vacuum. From the paraxial approximation follows that the electric field envelope $\vec{\widetilde{E}}_\bot = \vec{\bar{E}}_\bot \exp{[-i\omega z/c]}$ does not vary much along $z$ on the scale of the reduced wavelength $\lambda/(2\pi)$. As a result, the following field equation holds:

\begin{eqnarray} \mathcal{D}
\left[\vec{\widetilde{E}}_\bot(z,\vec{r}_\bot,\omega)\right] =
\vec{g}(z, \vec{r}_\bot,\omega) ~,\label{field1}
\end{eqnarray}
where the differential operator $\mathcal{D}$ is defined by

\begin{eqnarray}
\mathcal{D} \equiv \left({\nabla_\bot}^2 + {2 i \omega \over{c}}
{\partial\over{\partial z}}\right) ~,\label{Oop}
\end{eqnarray}
${\nabla_\bot}^2$ being the Laplacian operator over transverse cartesian coordinates. Eq. (\ref{field1}) is Maxwell's equation in paraxial approximation. The source-term vector $\vec{g}(z,\vec{r}_\bot)$ is specified by the trajectory of the source electrons, and can be written in terms of the Fourier transform of
the transverse current density, $\vec{\bar{j}}_\bot(z,\vec{r}_\bot,\omega)$, and of the charge density,
$\bar{\rho}(z,\vec{r}_\bot,\omega)$, as

\begin{eqnarray}
\vec{g} = && - {4 \pi} \left(\frac{i\omega}{c^2}\vec{\bar{j}}_\bot
-\vec{\nabla}_\bot \bar{\rho}\right) \exp\left[-\frac{i \omega
z}{c}\right] ~. \label{fv}
\end{eqnarray}
In this paper we will treat $\vec{\bar{j}}_\bot$ and $\bar{\rho}$ as macroscopic quantities, without investigating individual electron contributions. $\vec{\bar{j}}_\bot$ and $\bar{\rho}$ are regarded as given data, that can be obtained from any FEL code. Codes actually provide the charge density of the modulated
electron beam in the time domain $\rho(z,\vec{r}_\bot,t)$. A post-processor can then be used in order to perform the Fourier transform of $\rho$, that can always be presented as

\begin{eqnarray}
\bar{\rho} = \widetilde{\rho}(z,\vec{r}_\bot-\vec{r}_{o
\bot}(z),\omega) \exp\left[
i\omega\frac{s_o(z)}{v_o}\right]~,\label{rhotr}
\end{eqnarray}
where $\vec{r}_{o \bot}(z)$, $s_o(z)$ and $v_o$ are the transverse position, the curvilinear abscissa and the velocity of a reference electron with nominal Lorentz factor $\gamma_o$ that is injected on axis with no deflection and is guided by the planar undulator field. Such electron follows a trajectory specified by $\vec{r}_{o\bot}(z)=r_{ox} \vec{e}_x+r_{oy} \vec{e}_y$ with

\begin{eqnarray}
&&r_{ox}(z) = \frac{ K }{\gamma_o k_w} \cos(k_w z)  ~~,~~~
r_{oy}(z) = 0 ~. \label{rhel0}
\end{eqnarray}
The corresponding velocity is described by $\vec{v}_{o\bot}(z)=v_{ox}\vec{e}_x+v_{oy}\vec{e}_y$ with

\begin{eqnarray}
&&v_{ox}(z) = -  \frac{ K c}{\gamma_o} \sin(k_w z) ~~,~~~v_{oy}(z)
= ~~0~. \label{vhel0}
\end{eqnarray}
In Eq. (\ref{rhel0}), $K=\lambda_w e H_w/(2\pi m_e c^2)$ is the undulator parameter, $\lambda_w = 2\pi/k_w$ being the undulator period, $(-e)$ the negative electron charge\footnote{The minus sign on the right hand side of Eq. (\ref{rhotr})  is introduced for notational convenience and in agreement with the minus sign of the electron charge.}, $H_w$ the maximal modulus of the undulator magnetic field on-axis, and $m_e$ the rest mass of the electron. In writing Eq. (\ref{rhotr}) we are assuming that $\bar{\rho}$ varies approximately as $\exp\left[i\omega {s_o(z)}/{v_o}\right]$ over a distance $z$ of order of several undulator periods.

We note that for a generic motion one has

\begin{equation}
\omega \left({s_o(z_2)-s_o(z_1)\over{v}}-{z_2-z_1\over{c}}\right) =
\int_{z_1}^{z_2} d \bar{z} \frac{\omega}{2 c\gamma_z^2(\bar{z})}~, \label{moregen}
\end{equation}
where $\gamma_z(z) = 1/\sqrt{1-v_{oz}(z)^2/c^2}$ and $v_{oz}(z) = \sqrt{v_o^2-v_{o\bot}(z)^2}$. Thus, with the help of Eq. (\ref{rhotr}), Eq. (\ref{fv}) can be presented as

\begin{eqnarray}
\vec{g} = && - {4 \pi} \exp\left[i \int_{0}^{z} d \bar{z} \frac{
\omega }{2 {\gamma}_z^2(\bar{z})  c}\right]
\left[\frac{i\omega}{c^2}\vec{v}_{o\bot}(z) -\vec{\nabla}_\bot
\right]\widetilde{\rho}(z,\vec{r}_\bot-\vec{r}_{o\bot}(z),\omega)
~,\cr && \label{fvtf}
\end{eqnarray}
where we used the fact that $\vec{\bar{j}}_\bot = \vec{v}_{o\bot}\bar{\rho}$. In fact, for each particle in the beam the relative deviation of the particles energy from $\gamma_o m_e c^2$
is small, i.e. $\delta\gamma/\gamma_o \ll 1$. Therefore we can neglect differences between the average transverse velocity of electrons $\langle\vec{v}_\bot \rangle$ and $\vec{v}_{o\bot}$, so that
$\vec{\bar{j}}_\bot \equiv \langle \vec{v}_\bot \rangle \bar{\rho} \simeq \vec{v}_{o\bot} \bar{\rho}$. Because of this we will also drop the subscript "$o$" in $\gamma_o$.

With the help of Eq. (\ref{fvtf}),  Eq. (\ref{field1}) can also be presented as:

\begin{eqnarray}
&&\left({\nabla}^2_\bot + \frac{2 i \omega}{c}
\frac{\partial}{\partial z}\right) \vec{\widetilde{E}}_\bot(z,\vec{r}_\bot,\omega) = \cr &&
-{4 \pi} \exp\left[i \int_{0}^{z} d \bar{z} \frac{
\omega }{2 c{\gamma}_z^2(\bar{z}) }\right] \left[\frac{i\omega}{c^2}\vec{v}_{o\bot}
-\vec{\nabla}_\bot \right] \tilde{\rho}(z,\vec{r}_\bot-\vec{r}_{o\bot}(z),\omega) ~, \label{incipit4}
\end{eqnarray}
where we have set

\begin{equation}
\int_{0}^{z} d \bar{z} \frac{
\omega }{2 c {\gamma}_z^2(\bar{z})}  \simeq \frac{K^2}{4\gamma^2 - K^2 } \frac{\omega z}{c}  - \frac{\omega K^2}{8 c \gamma^2
k_w}\sin\left(2 k_w z \right). \label{phu}
\end{equation}
With the aid of the appropriate Green's function an exact solution of Eq. (\ref{incipit4}) can be found without any extra assumption about the parameters of the problem:

\begin{eqnarray}
&&\vec{\widetilde{E}}_\bot(z, \vec{r}_\bot, \omega )=
\int_{-\infty}^{\infty} dz' \frac{1}{z-z'} \int d
\vec{r'}_\bot \left[\frac{i\omega}{c^2}\vec{v}_{o\bot}(z')
-\vec{\nabla}'_\bot \right]\cr &&\times
\tilde{\rho}(z',\vec{r'}_\bot-\vec{r}_{o\bot}(z'),\omega)
\exp\left\{i\omega\left[\frac{\mid \vec{r}_\bot-\vec{r'}_\bot
\mid^2}{2c (z-z')}\right]+ i \left[\int_{0}^{z'} d \bar{z} \frac{
\omega }{2c {\gamma}_z^2(\bar{z}) }\right] \right\} ~,\cr && \label{blob}
\end{eqnarray}
where $\vec{\nabla}'_\bot$ represents the gradient operator with respect to the source point, while $(z, \vec{r}_\bot)$ indicates the observation point. Integration by parts of the gradient terms leads to

\begin{eqnarray}
&&\vec{\widetilde{E}}_\bot(z, \vec{r}_\bot,\omega )= \frac{i \omega }{c}
\int_{-\infty}^{\infty} dz' \frac{1}{z-z'}  \int d
\vec{r'}_\bot \left(\frac{\vec{v}_{o\bot}(z')}{c}
-\frac{\vec{r}_\bot-\vec{r'}_\bot}{z-z'}\right)\cr&&\times
\tilde{\rho}(z',\vec{r'}_\bot-\vec{r}_{o\bot}(z'),\omega)\exp\left[i
\Phi_T(z',\vec{r'}_\bot,\omega)\right] ~, \cr &&
\label{generalfin}
\end{eqnarray}
where the total phase $\Phi_T$ is given by

\begin{equation}
\Phi_T =  \left[\int_{0}^{z'} d \bar{z} \frac{
\omega }{2 c {\gamma}_z^2(\bar{z})  }\right]+ \omega \left[
\frac{|\vec{r}_\bot-\vec{r'}_\bot|^2}{2c (z-z')}\right]~.
\label{totph}
\end{equation}
We now make use of a new integration variable $\vec{l}=\vec{r'}_\bot-\vec{r}_{o\bot}(z')$ so that

\begin{eqnarray}
&&\vec{\widetilde{E}}_\bot(z, \vec{r}_\bot, \omega )= \frac{i \omega }{c}
\int_{-\infty}^{\infty} dz' \frac{1}{z-z'}  \int d \vec{l}
\left(\frac{\vec{v}_{o\bot}(z')}{c} -\frac{\vec{r}_\bot-\vec{r}_{o\bot}(z')-\vec{l}}{z-z'}\right)\cr&&
\times \tilde{\rho}(z',\vec{l},\omega) \exp\left[i \Phi_T(z',\vec{l},\omega)\right] ~,
\label{generalfin2}
\end{eqnarray}
and

\begin{equation}
\Phi_T =  \left[\int_{0}^{z'} d \bar{z} \frac{
\omega }{2 c{\gamma}_z^2(\bar{z})  }\right]+ \omega \left[
\frac{|\vec{r}_\bot-\vec{r}_{o\bot}(z')-\vec{l}|^2}{2c
(z-z')}\right] ~. \label{totph2}
\end{equation}

Introducing the far zone approximation and making the appropriate substitutions into Eq. (\ref{generalfin2}) yields the following field contribution calculated along the undulator:

\begin{eqnarray}
\vec{\widetilde{E}}_\bot(z, \vec{r}_\bot, \omega ) &=& -\frac{i \omega }{c z} \int
d\vec{l} \int_{-\infty}^{\infty} dz' \tilde{\rho}(z',\vec{l},\omega) {\exp{\left[i \Phi_T(z',\vec{l},\omega)\right]}}
\cr &&\times \left[\left(\frac{K}{\gamma} \sin\left(k_w
z'\right)+\theta_x\right)\vec{e}_x
+\theta_y \vec{e}_y\right]~, \label{undurad}
\end{eqnarray}
where

\begin{eqnarray}
\Phi_T &=& {\omega} \left\{ \frac{z'}{2\gamma^2 c}
\left[1+\frac{K^2}{2} +
\gamma^2\left(\theta_x^2+\theta_y^2\right)\right]\right.
\cr&&\left.-\frac{K^2}{8\gamma^2k_w c} \sin{(2 k_w z')} - \frac{K
\theta_x}{\gamma k_w c} \cos{(k_w z')}\right\} \cr&&
+\omega \left\{\frac{K \theta_x}{k_w\gamma c}
-\frac{1}{c}(\theta_x l_x+ \theta_y l_y) + (\theta_x^2+\theta_y^2)
\frac{z}{2c} \right\} ~.\label{phitundu}
\end{eqnarray}
Here $\theta_{x}$ and $\theta_{y}$ indicate the observation angles $x/z$ and $y/z$. Moreover, since in Eq. (\ref{undurad}) we introduced explicitly the trajectory inside the undulator, we need to limit the integration in $dz'$ to a proper range within the undulator. We assume that this is done by introducing a proper function of $z'$ as a factor $\widetilde{\rho}$, which becomes zero outside properly defined range, thus effectively limiting the integration range in $z'$.

In this article we are interested in considering fields and electromagnetic sources originating from an FEL process. Imposing resonance condition between electric field and reference particle, the self-consistent FEL process automatically restricts the amplification of radiation at frequencies around the first harmonic:

\begin{equation}
\omega_{1o} = 2 {k_w c \bar{\gamma}_z^2} ~,\label{freqfix}
\end{equation}
where

\begin{equation}
\bar{\gamma}_z^2 = \frac{\gamma^2}{1+K^2/2}~, \label{gammaz}
\end{equation}
and at emission angles $\theta_\mathrm{max}^2 \ll 1/\bar{\gamma}_z^2$. Our focus onto FEL emission also explains the definition in Eq. (\ref{rhotr}). In fact, introduction of $\widetilde{\rho}$ is useful when $\widetilde{\rho}$ is a slowly varying function of $z$ on the wavelength scale. If the charge density distribution under study originates from an FEL process a stronger condition is satisfied, namely $\widetilde{\rho}$ is slowly varying on the scale of the undulator period $\lambda_w$ and, as the FEL pulse itself, is peaked around the fundamental $\omega_{1o}$. The words "peaked" or "around" the fundamental mean that the bandwidth is $\Delta \omega/\omega_{1o} \ll 1$. We quantify "how near" the frequency $\omega$ is to $\omega_{1o}$ introducing the detuning parameter $C$, defined as

\begin{eqnarray}
C = \frac{\omega}{2c \bar{\gamma}_z^2 } -  k_w = \frac{\Delta
\omega}{\omega_{1o}} k_w~ \label{Ch}
\end{eqnarray}
with $\Delta \omega = \omega -  \omega_{1o}$. The detuning parameter $C$ should indeed be considered as a function of $z$, $C=C(z)$ because, in general, the parameters of the undulator my be a function of $z$. All other dependencies on $z$, for example due to the fact that the energy of particles actually deviates from $\gamma$ and actually decelerate during the FEL process, is accounted for in $\tilde{\rho}$. Indicating with $\vec{\widetilde{E}}_{\bot1}$, we seek to calculate the first harmonic contribution at frequencies $\omega$ around $\omega_{1o}$ by making use of the well-known expansion

\begin{equation}
\exp{[i a \sin{(\psi)}]}=\sum_{p=-\infty}^{\infty} J_p(a) \exp{[i
p \psi]}~, \label{alfeq} \end{equation}
where $J_p$ indicates the Bessel function of the first kind of order $n$. This allows to cast Eq. (\ref{undurad}) into a form that is still fully general, but more convenient for enforcing the conditions, discussed above, related with FEL emission. One obtains

\begin{eqnarray}
\vec{\widetilde{E}}_{ 1}&=& -\frac{i \omega_{1o}  }{c z}
\int_{-\infty}^{\infty} d l_x \int_{-\infty}^{\infty} d l_y
\int_{-\infty}^{\infty} dz' \tilde{\rho}(z',\vec{l},\omega)
\left(z',\omega \right) \exp[i\Phi_o] \cr &&\times
\sum_{m=-\infty}^{\infty} \sum_{n=-\infty}^{\infty} J_m(u)
J_{n}(v)  \exp\left[\frac{i n \pi}{2}\right] \cr && \times
\Bigg\{\Bigg[-i \frac{K}{2\gamma}
\Bigg(\exp\left\{i[R_\omega+1]k_w z'\right\}
-\exp\left\{i[R_\omega-1]k_w z'\right\}\Bigg)
\cr&&+\theta_x \exp\left\{i R_\omega k_w z'\right\}
\Bigg]\vec{e}_x +\Bigg[\theta_y \exp\left\{i R_\omega k_w
z'\right\} \Bigg]\vec{e}_y\Bigg\} ~, \label{undurad2}
\end{eqnarray}
where

\begin{equation}
R_\omega = \frac{\omega}{\omega_1} - n -2 m ~,\label{romega}
\end{equation}
with

\begin{equation}
\frac{1}{\omega_1}=
\frac{1}{\omega_{1o}}\left[1+\bar{\gamma}_z^2\left(
\theta_x^2+
\theta_y^2\right)\right]\label{omega1}~.
\end{equation}
Moreover

\begin{equation}
u=\frac{\omega_{1o}}{\omega_1} \frac{K^2
\left[1-K^2/(4\gamma^2)\right]}
{4\left[1+{K^2}/{2}+\gamma^2\left(
\theta_x^2+
\theta_y^2\right)\right]} \simeq  \frac{K^2}{2(2+K^2)}
\label{v},~
\end{equation}
\begin{equation}
v=\frac{\omega_{1o}}{\omega_1} \frac{2K  \gamma
\left[1-K^2/(4\gamma^2)\right]\theta_x}
{1+{K^2}/{2}+\gamma^2\left(\theta_x^2+
\theta_y^2\right)} \simeq \frac{4 K  \gamma
\theta_x}{2+K^2} \label{u}~
\end{equation}
and
\begin{equation}
\Phi_o = \omega_{1o} \left[\frac{K \theta_x}{k_w\gamma c}
- \frac{1}{c}(\theta_x l_x+\theta_y
l_y)+\frac{z}{2c}(\theta_x^2+\theta_y^2) \right]~. \label{phio}
\end{equation}
Since the FEL process imposes that the magnitude of the charge density $\tilde{\rho}$ does not vary much in $z'$ over a period of the undulator $\lambda_w$, it follows that the fast oscillations in the exponential function in the integrand of Eq. (\ref{undurad2}) tend to suppress the integral unless $R_\omega = 0$, $R_\omega=-1$  or $R_\omega = 1$, that is when at least one of the exponential function is simply unity. Using Eq. (\ref{Ch}) and Eq. (\ref{omega1})we rewrite Eq. (\ref{romega}) as

\begin{equation}
R_\omega = \left(1+\frac{C}{k_w}\right)\left[1+\bar{\gamma}_z^2\left(
\theta_x^2+
\theta_y^2\right)\right] - n -2 m ~,\label{romega2}
\end{equation}
Invoking again the FEL process, inspection of Eq. (\ref{romega2}) shows that in the limit for $C \ll k_w$ and $\theta_\mathrm{max}^2 \ll 1/\bar{\gamma}_z^2$  the conditions above correspond to  $n = 1 - 2m$, $n=2-2 m $ and $n=-2 m $ respectively.  Neglecting all other terms and imposing $\omega = \omega_1 + \Delta \omega_1$ we obtain

\begin{eqnarray}
\vec{\widetilde{E}}_{\bot1} &=& -\frac{i \omega_{1o}  }{c z}
\int_{-\infty}^{\infty} d l_x \int_{-\infty}^{\infty} d l_y
\int_{-\infty}^{\infty} dz' \tilde{\rho}(z',\vec{l},\omega) \exp[i\Phi_o] \exp\left\{i\frac{\Delta
\omega_1}{\omega_1} k_w z'\right\}  \cr &&
\times\sum_{m=-\infty}^{\infty} \Bigg\{\Bigg[-i \frac{K}{2\gamma}
\left( J_m(u) J_{2-2m}(v) \exp\left\{\frac{i [2-2m]
\pi}{2}\right\} \right.\cr&&\left. -J_m(u)  J_{-2m}(v)
\exp\left\{\frac{i [-2m] \pi}{2}\right\}\right) \cr&&
+\theta_x J_m(u) J_{1-2m}(v) \exp\left\{\frac{i [1-2m]
\pi}{2}\right\} \Bigg]\vec{e}_x \cr&&+\Bigg[\theta_y J_m(u)
J_{1-2m}(v) \exp\left\{\frac{i [1-2m]\pi}{2}\right\}
\Bigg]\vec{e}_y\Bigg\} ~. \label{undurad3}
\end{eqnarray}
For any value of $K$ and $\theta_{x}$ much smaller than $1/\bar{\gamma}_z$, $v$ is a small parameter. This allows one further mathematical step, that is the expansion of the Bessels functions $J_q(v)\sim v^{|q|}$ in Eq. (\ref{undurad2}). If one retains only the smallest indexes, one obtains the usual expression for the first harmonic of the undulator field under the resonance approximation, which is linearly polarized along the $\vec{e}_x$ direction. This can be done by keeping the current term with $m=1$ (that is the first of the two terms in the $\vec{e}_x$ direction), and neglecting everything else. However, here we want to investigate corrections to the field along the $\vec{e}_y$ direction. With this in mind, we also retain gradient terms corresponding to $m=0$ and $m=-1$, which correspond to the largest corrections along the $\vec{e}_y$  Eq. (\ref{undurad3}) then yields:

\begin{eqnarray}
\vec{\widetilde{E}}_{\bot1}&=&\frac{ \omega_{1o} }{c z}
\left[\frac{K}{2\gamma}\mathrm{A_{JJ}} \vec{e}_x +   \frac{2 K \gamma}{2+K^2}\mathrm{B_{JJ}} \theta_x \theta_y \vec{e}_y\right] \int_{-\infty}^{\infty}
d l_x \int_{-\infty}^{\infty} d l_y \int_{-\infty}^{\infty} d z'
\exp{[i\Phi_o]} \cr && \times \tilde{\rho}(z',\vec{l},\omega) \exp[i C z']  \exp\left[i
\bar{\gamma}_z^2 \left(\theta_x^2+
\theta_y^2\right) k_w z'\right]
\label{undurad4}
\end{eqnarray}
where we have defined

\begin{equation}
\mathrm{A_{JJ}} = J_0\left(\frac{K^2}{2(2+K^2)}\right)-J_1\left(\frac{K^2}{2(2+K^2)}\right)
~, \label{calA}
\end{equation}
\begin{equation}
\mathrm{B_{JJ}}  = J_0\left(\frac{K^2}{2(2+K^2)}\right)+J_1\left(\frac{K^2}{2(2+K^2)}\right)~, \label{calB}
\end{equation}
and

\begin{equation}
\Phi_o = \frac{\omega_{1o}}{c} \left[ - (\theta_x l_x+\theta_y
l_y)+\frac{z}{2}(\theta_x^2+\theta_y^2) \right]~.
\label{phiosimpli}
\end{equation}
Eq. (\ref{undurad4}) can be also written as:

\begin{eqnarray}
\vec{\widetilde{E}}_{\bot1}&=&\frac{ \omega_{1o}} {c z}
\exp{\left[i \frac{\omega_{1o}}{2c}
{z}({\theta}_x^2+{\theta}_y^2) \right]}
\left[\frac{K}{2\gamma}\mathrm{A_{JJ}}\vec{e}_x +
\frac{2 K \gamma}{2+K^2}\mathrm{B_{JJ}}{\theta}_x{\theta}_y\vec{e}_y\right]\cr&&
\times\int_{-\infty}^{\infty} d {l}_x \int_{-\infty}^{\infty} d
{l}_y \int_{-\infty}^{\infty} d {z}' \exp{\left[-i
\frac{\omega_{1o}}{c} \left({\theta}_x l_x + \theta_y
l_y\right)\right] } \cr &&\times \exp\left[i \frac{\omega_{1o}}{2
c} \left(\theta_x^2+\theta_y^2\right) z'\right]\tilde{\rho}(z',\vec{l},\omega) \exp[i C z']   ~,\label{undurad5}
\end{eqnarray}
Note that usually computer codes present the product $\tilde{\rho}(z',\vec{l},\omega) \exp[i C z']$ combined in a single quantity tipically known as the complex amplitude of the electron beam modulation with respect to the phase $\psi = k_w z' + (\omega/c) z' - \omega t$. Regarding such product as a given function allows one not to bother about a particular presentation of the beam modulation. Eq. (\ref{undurad4}) or, equivalently, Eq. (\ref{undurad5}) are our most general result, and are valid independently on the model chosen for the current density and the modulation and can be used together with FEL simulation codes for detailed calculations of the evolution of the vertically polarization contribution to the FEL radiation.

\subsection{Discussion: other contributions to the vertically polarized field}

Inspection of Eq. (\ref{undurad4}) shows that the vertical polarization component depends on angles and roughly scales, aside for the $K$-dependent correction factor $B_\mathrm{JJ}$ of order unity, as $2 K \bar{\gamma}_z \theta_x \theta_y$. Therefore, the ratio between vertical and horizontal polarization components scales as $2 \bar{\gamma}_z^2 \theta_x \theta_y$. It is useful to underline that the vertical-polarization correction was calculated on the basis of the leading resonant contribution to the vertically polarized field. A natural question pertains the magnitude of the non-resonant corrections, which -roughly speaking- are expected to be suppressed due to fast oscillations in the exponential function in the integrand of Eq. (\ref{undurad2}). The leading non-resonant correction term is found in Eq. (\ref{undurad2}) setting $R_\omega = \pm 1$ and choosing $n=0$, which is justified because $v \ll 1$. The oscillating integrand in $d z'$ suppresses the correction term of a factor proportional to the inverse number of undulator periods involved in the FEL process, that is those in a field gain length, $N_w^{-1}$, and the correction term is therefore proportional to $\theta_y/N_w$ aside, again, for the $K$-dependent correction factor $B_\mathrm{JJ}$ of order unity. Therefore, altogether, the ratio between the leading non-resonant correction term and the horizontal polarization component scales as $\bar{\gamma}_z \theta_y/N_w$. It follows from the ratio with the horizontally polarized main component of the radiation that the non-resonant correction terms can be neglected whenever $N_w \bar{\gamma}_z \theta_x \gg 1$. This means, for example, that on-axis at $\theta_x = 0$ the first non-resonant term is larger than the first resonant term, and seems in contradiction with our choice to neglect non-resonant terms. However, here we are interested in the different polarization contributions to the angle-integrated powers, and not to the intensity at a fixed angle. As already discussed, the presence of the FEL process limits the detuning to values $C \ll k_w$ and the angles of interest up to $\theta_\mathrm{max}^2 \sim 1/(N_w \bar{\gamma}_z^2) \ll 1/\bar{\gamma}_z^2$. From Eq. (\ref{undurad4}) follows that the contribution from the leading resonant term is of order $4 \bar{\gamma}_z^4 \int_0^{2\pi} d\phi\int_0^{\theta_\mathrm{max}} d\theta ~ \theta^5 \cos^2(\phi) \sin^2(\phi) = \pi /(6 \bar{\gamma}_z^{2} N_w^{3})$, while from Eq. (\ref{undurad2}) one obtains that the contribution of the leading non-resonant term is of order $\bar{\gamma}_z^2 N_w^{-2} \int_0^{2\pi} d\phi\int_0^{\theta_\mathrm{max}} d\theta ~ \theta^3 \sin^2(\phi) =\pi /(4 \bar{\gamma}_z^{2} N_w^{4})$. Therefore, the non-resonant contribution to the angle-integrated polarization correction is fully negligible, because the ratio between resonant and non-resonant contributions scales as $N_w \gg 1$. Similar reasoning can be made, starting from Eq. (\ref{generalfin2}), for the case of any other non-resonant contribution arising from any part of the electron trajectory, not only the undulator. The specific angular dependence will vary from the kind of contribution but in general one will obtain a field scaling as $1/N_w$ because it is non-resonant. The transverse distribution extends over angles of order $\sqrt{N_w}$ times larger than $\theta_\mathrm{max}$ for the same reason. Therefore, one can generalize the argument above, saying that, as long as one consider an angular acceptance of order $\theta_\mathrm{max}$, the power contribution due to any non-resonant terms will scale as $1/N_w^4$ and will be negligible.

Summing up, in the case of an FEL, due to the presence of a maximum angle $\theta_\mathrm{max}$ related with the self-consistent process, the angle-integrated correction to the power from the horizontally polarized radiation component only includes the leading resonant term, and Eq. (\ref{undurad4}) can always be used to calculate such correction at the first harmonic.
%
%At variance, the situation becomes much more complicated in the case one is interested in spontaneous emission from an undulator, without an FEL process. In fact, in that case one does not have a maximal angle $\theta_\mathrm{max}$ limited by the presence of the FEL self-consistent process. In the case of spontaneous emission, the angle $\theta_\mathrm{max}$ measures the aperture of the central cone only. It is true that the flux of undulator radiation is concentrated within such cone, while non-resonant contributions are spread over a much larger angle. However, the angle-integrated fluxes are comparable. Therefore, there is a fundamental difference between the polarization correction in the case of FEL emission and of undulator emission. In the case of undulator emission the ratio of the polarization contributions heavily depends on the aperture slits over which radiation is collected and in particular, integrating over all angles the correction to the horizontal polarization component must include non-resonant terms. As a result, our expression can be used for the spontaneous emission from undulators  only in some cases, for example when slits limit the collection of radiation to an angle $\theta_\mathrm{max}$. In the case of an FEL, due to the presence of a maximum angle $\theta_\mathrm{max}$ related with the self-consistent process, the angle-integrated correction to the power from the horizontally polarized radiation component only includes the leading resonant term, and Eq. (\ref{undurad4}) can always be used to calculate such correction.

\section{\label{sec:mod} Physical situations treatable analytically}

In the previous section we calculated the two main different polarization contributions to the radiation from an FEL with a planar undulator. One originates from the horizontal component $\widetilde{E}_x$ of the electric field and the other from the vertical component $\widetilde{E}_y$, respectively the $\sigma$-mode and the $\pi$-mode.

We now restrict our attention to the steady-state model of an FEL amplifier. As we will see in the first part of this section, the simplicity of the steady state model offers the opportunity for an almost complete analytical description of the polarization characteristics of an FEL amplifier in the high-gain linear mode of operation. The second part of the section is devoted, instead, to an analytical solution for the case of a constant electron density modulation. This model is important from methodological point of view, since it allows one to imitate the polarization characteristics of FEL radiation at saturation. These analytical solutions can serve as a reliable basis for the development of numerical methods.

Because of the steady state   assumption we restrict our attention to one single frequency. This means that, in the time domain, the electric field envelope $\vec{\widetilde{E}}_{\bot1}$ must correspond to a real electric field at a certain frequency $\bar{\omega} = \omega_{1o} (1 + C k_w)$ given by $\vec{E}(z,\vec{r}_\bot,t) = {\vec{\mathcal{E}}}_{\bot 1}(z,\vec{r}_\bot) \exp[i \bar{\omega} (z/c - t)] + C.C.$, where the symbol $C.C.$ indicates complex conjugation. In the following we will be interested in integrating the angular power distribution of radiation to obtain the power fractions into the two modes of polarization. The power for the $\sigma$- and $\pi$-polarization components of the first harmonic radiation can be found remembering that the Poynting vector $\vec{S} = c/(4\pi) \vec{E} \times \vec{B}$ represents the energy flow per unit time and per unit area. The average power carried by an electromagnetic signal per unit area can be found by averaging over the signal duration. For a monochromatic signal, we further need to take a limit for an infinitely long signal. Distinguishing between the two polarization modes this procedure amounts to

\begin{eqnarray}
\frac{dW_{(\sigma,\pi)}}{dS} = \frac{c}{4\pi} \lim\limits_{T\rightarrow \infty} \frac{1}{T} \int_{-T/2}^{T/2} dt  \left|{E}_{(x,y)}(z,\vec{r}_\bot,t) \right|^2 ~,
\label{unoo}
\end{eqnarray}
where we used the fact that in the paraxial approximation and in cgs units $|\vec{E}|=|\vec{B}|$. From our previous discussion we can write

\begin{eqnarray}
\left|{E}_{(x,y)}(z,\vec{r}_\bot,t) \right|^2 = 4 \left|\mathcal{E}_{\bot1 (x,y)}(z,\vec{r}_\bot) \right|^2\cos^2[\omega_{1o} t+\phi(z)]~.
\label{interm1}
\end{eqnarray}
Substitution in Eq. (\ref{unoo}) yields

\begin{eqnarray}
&&\frac{dW_{(\sigma,\pi)}}{dS} = \frac{c}{\pi}\left|\mathcal{E}_{\bot1 (x,y)}(z,\vec{r}_\bot) \right|^2 \lim\limits_{T\rightarrow \infty} \frac{1}{T}  \int_{-T/2}^{T/2} dt  \cos^2[\omega_{1o} t+\phi(z)]  \cr &&  = \frac{c}{2\pi} \left|\mathcal{E}_{\bot 1 (x,y)}(z,\vec{r}_\bot) \right|^2 ~.
\label{unoob}
\end{eqnarray}
Finally, integrating over $dS$, one obtains the average power for the two fractions

\begin{eqnarray}
W_{(\sigma,\pi)} &=&  \frac{c}{2 \pi} \int_{-\infty}^{\infty} dx
\int_{-\infty}^{\infty} dy {|\mathcal{E}_{\bot1 (x,y)}(z, x,
y)|^2}~, \label{xpowden}
\end{eqnarray}
In order to find $\mathcal{E}_{\bot1 (x,y)}$ we note that the Fourier transform of our real monochromatic field is given by

\begin{eqnarray}
&&\vec{\bar{E}}_{\bot}(z,\vec{r}_\bot,\omega)  \cr && =
2 \pi \exp\left(i \frac{{\omega}}{c} z\right)  \left[ {\vec{\mathcal{E}}}_{\bot 1}(z,\vec{r}_\bot) \delta(\omega-\bar{\omega}) +   {\vec{\mathcal{E}}}_{\bot 1}^*(z,\vec{r}_\bot)  \delta(\omega+\bar{\omega}) \right]~,
\label{FTfieldom0}
\end{eqnarray}
where the last equality follows from direct application of the Fourier transform shift theorem, since the dependence of the monochromatic field on time is $z/c - t$. By definition of field envelope one has thus

\begin{eqnarray}
\vec{\widetilde{E}}_{\bot1}(z,\vec{r}_\bot,\omega)=2 \pi \left[ {\vec{\mathcal{E}}}_{\bot 1}(z,\vec{r}_\bot) \delta(\omega-\bar{\omega}) +   {\vec{\mathcal{E}}}_{\bot 1}^*(z,\vec{r}_\bot)  \delta(\omega+\bar{\omega}) \right]~.
\label{eccola0}
\end{eqnarray}
Passing to the complex notation for the field in the time domain (i.e. dropping the complex conjugate term), one can simplify the previous equation keeping only the positive frequency $\bar{\omega}$ and write

\begin{eqnarray}
\vec{\widetilde{E}}_{\bot1}(z,\vec{r}_\bot,\omega)=2 \pi {\vec{\mathcal{E}}}_{\bot 1}(z,\vec{r}_\bot) \delta(\omega-\bar{\omega}) ~.
\label{eccola}
\end{eqnarray}
Summing up, in order to calculate $W_{(\sigma,\pi)}$ we should make use of Eq. (\ref{xpowden}). In order to do so, we need an expression for $\mathcal{E}_{\bot1 (x,y)}$, which can be found in terms of $\vec{\widetilde{E}}_{\bot1}$ with the help of Eq. (\ref{eccola}). Finally, one needs to calculate $\vec{\widetilde{E}}_{\bot1}$, which can be done using Eq. (\ref{undurad5}).

Let us therefore turn to the calculation of $\vec{\widetilde{E}}_{\bot1}$ in Eq. (\ref{undurad5}), which requires specification of the bunching $\tilde{\rho}$. Formally,  the one-dimensional steady state theory of FEL amplifiers  deals with the amplification of a plane electromagnetic wave by an infinitely wide and infinitely long electron beam. The approximation of an infinitely long electron beam is acceptable when one considers a uniformly dense beam that is much longer than the slippage in a gain length. However, even in this case the electron beam and the electromagnetic wave have finite transverse dimensions, and diffraction effects always take place. For the one-dimensional model to be applicable diffraction losses must be negligible. In practice, such assumption is valid for XFELs operating in the hard X-ray wavelength range. In this case, the asymptote for the field growth rate of the fundamental $\mathrm{TEM_{00}}$ mode can be found from the cubic eigenvalue equation of the one-dimensional model of FEL amplifier. We also assume that the transverse current density of the electron beam can be modeled as a Gaussian.

With these assumptions, we may write the charge density $\rho(z,\vec{r}_\bot,t)$ in the time domain as

\begin{eqnarray}
\rho(z,\vec{r}_\bot,t) =  \frac{1}{v_z} \rho_\bot(\vec{r}_\bot-\vec{r}_{o\bot}) f\left(z,t-\frac{s_o}{v_o}\right)~.
\label{rhobotime}
\end{eqnarray}
The quantity $\rho_\bot$ has the meaning of transverse charge density distribution, while $f$ is the temporal electron number density distribution. If the bunch is modulated at a given frequency $\bar{\omega}$, one has

\begin{eqnarray}
f\left(z,t-\frac{s_o}{v_o}\right) = f_0\left(t-\frac{s_o}{v_o}\right) \left(1+ \left\{a(z) \exp\left[i \bar{\omega} \left(t-\frac{s_o}{v_o}\right) \right] + C.C.\right\}\right)
\label{defi}
\end{eqnarray}
where $f_0$ is the temporal charge density distribution of an unmodulated bunch, assumed to be constant along the undulator, while $a$ is the modulation or bunching amplitude. Note that in the general case, the bunching amplitude is a complex function of the position in the undulator $z$.

We start the analysis for the case of a stepped profile electron pulse of finite duration $T$ and then we go over to the limit of an infinitely long pulse. The  beam current can be written in the form:

\begin{eqnarray}
I= \frac{(-e) N}{T} H_T\left(t-\frac{s_o}{v_o}\right) \left(1+ \left\{a(z) \exp\left[i \bar{\omega} \left(t-\frac{s_o}{v_o}\right) \right] + C.C.\right\}\right)~,\cr &&
\label{Ibeamt}
\end{eqnarray}
where $H_T(t) = 1$ in the range $(-T/2,T/2)$ and zero otherwise, $N$ is the number of electrons in the pulse and $(-e)N/T = - I_0$ is the average beam current. Comparison with Eq. (\ref{defi}) yields

\begin{eqnarray}
f = \frac{1}{T} H_T\left(t-\frac{s_o}{v_o}\right) \left(1+ \left\{a(z) \exp\left[i \bar{\omega} \left(z,t-\frac{s_o}{v_o}\right) \right] + C.C.\right\}\right) ~.
\label{defi2}
\end{eqnarray}
Assuming a Gaussian transverse charge density distribution of the electron beam with rms size $\sigma$ given by

\begin{eqnarray}
\rho_\bot(\vec{r}_\bot) =  \frac{(-e) N}{2 \pi {\sigma^2}} \exp\left(-\frac{{r}_\bot^2}{2 \sigma^2}\right)
\label{trchdenst}
\end{eqnarray}
we thus obtain

\begin{eqnarray}
&&\rho(z,\vec{r}_\bot,t) = -\frac{1}{v_z} \frac{I_o}{2 \pi {\sigma^2}} \exp\left(-\frac{{r}_\bot^2}{2 \sigma^2}\right) H_T \left(z,t-\frac{s_o}{v_o}\right) \cr && \times \left(1+ \left\{a(z) \exp\left[i \bar{\omega}  \left(z,t-\frac{s_o}{v_o}\right)\right] + C.C.\right\}\right)~.
\label{rhoperp2time}
\end{eqnarray}
Let us study the asymptote for an infinitely long electron beam, that is the limit for $T \longrightarrow \infty$, $N \longrightarrow \infty$ and $(-e)N/T = \mathrm{constant} = -I_0$. One needs to calculate the temporal Fourier transform of $\rho$. Fourier-transforming Eq. (\ref{rhoperp2time}) and using Eq. (\ref{rhotr}) we write the expression for the slowly-varying amplitude

\begin{eqnarray}
\tilde{\rho}(z, \vec{r}_\bot, \omega) = \frac{j_o(\vec{r}_\bot)}{v_z}  2\pi a(z) \delta(\omega-\omega_0) ~,
\label{tildethop}
\end{eqnarray}
where we defined the current density

\begin{equation}
j_o(\vec{r}_\bot) = -\frac{I_o}{2 \pi \sigma^2} \exp\left(-\frac{r_\bot^2}{2 \sigma^2}\right) ~,
\label{exbot2}
\end{equation}
and where we dropped the term in $\delta(\omega+\bar{\omega})$ passing to complex notation, as done before with the field.

\subsection{High-gain linear regime}

We first model the case of an FEL amplifier in the high-gain linear regime.  We proceed   approximating the detuning parameter $C$  as constant along the undulator. Let us restrict, for simplicity, to the case of perfect resonance for $C=0$. This means that from now on $\bar{\omega} = \omega_{1o}$. The high-gain asymptote of the one-dimensional steady-state theory of FEL amplifiers yields

\begin{eqnarray}
a(z) = a_f \exp[(\sqrt{3}+i)z/(2 L_g)] ~,
\label{fzzz}
\end{eqnarray}
where we set the exit of the undulator (in the linear regime) at $z=0$ and  $a_f=\mathrm{constant}$ is the modulation level at $z=0$. Here $L_g$ is the field gain length. The number of undulator periods in the field gain length $L_g$ is just $N_w = (4 \pi \rho_\mathrm{1D})^{-1}$, where the FEL parameter $\rho_\mathrm{1D}$ \cite{BONI} is related to the problem parameters through

\begin{eqnarray}
\rho_\mathrm{1D} = \frac{\lambda_w}{4\pi} \left[\frac{\pi I_0 K^2 A_{JJ}^2}{I_A \sigma^2 \lambda_w \gamma^3}\right]^{1/3}~,
\label{rho1d}
\end{eqnarray}
where $I_A = m_e c^3/e \simeq 17$ kA is the Alfven current. If we now integrate Eq. (\ref{undurad5}) from $z = -\infty$ to $z=0$ we get

\begin{eqnarray}
\vec{\widetilde{E}}_{\bot 1}&=&\frac{2 \pi a_f \omega_{1o}} {c^2 z}
\exp{\left[i \frac{\omega_{1o}}{2c}
{z}({\theta}_x^2+{\theta}_y^2) \right]}
\left[\frac{K}{2\gamma}\mathrm{A_{JJ}}\vec{e}_x+
\frac{2 K \gamma}{2+K^2}\mathrm{B_{JJ}}{\theta}_x {\theta}_y \vec{e}_y\right]\cr&&
\times\int_{-\infty}^{\infty} d {l}_x \int_{-\infty}^{\infty} d
{l}_y \int_{-\infty}^{0} d {z}' \exp{\left[-i
\frac{\omega_{1o}}{c} \left({\theta}_x l_x +
\theta_y l_y\right)\right] } \cr &&\times
\exp\left[i \frac{\omega_{1o}}{2 c}
\left(\theta_x^2+ \theta_y^2\right) z'\right]
j_{o}\left(\vec{l}\right)  \exp\left[\frac{(\sqrt{3}+i)z'}{2 L_g}\right] \delta(\omega-\omega_{1o})
 ~,\label{undurad6new}
\end{eqnarray}
where we substituted $v_z$ with $c$ in Eq. (\ref{tildethop}), based on the fact that $1/\gamma_z^2 \ll 1$.

If we now substitute Eq. (\ref{exbot2}) in Eq. (\ref{undurad6new}) and we perform all integrations, Eq. (\ref{eccola}) yields

\begin{eqnarray}
\vec{\mathcal{E}}_{\bot 1}&=&-\frac{ 2 I_o a_{f}\omega_{1o} L_g}
{c^2 z} \exp{\left[i
\frac{\omega_{1o}}{2c}{z}({\theta}_x^2+{\theta}_y^2) \right]}
\left[\frac{K}{2\gamma}\mathrm{A_{JJ}}\vec{e}_x +
\frac{2 K \gamma}{2+K^2}\mathrm{B_{JJ}} {\theta}_x {\theta}_y\vec{e}_y\right] \cr&&
\times \frac{\exp{\left[-{\sigma^2 \omega_{1o}^2 } \left(
\theta_x^2 + \theta_y^2 \right)/({2 c^2})\right]}}{(\sqrt{3}+i) c + i L_g \omega_{1o} \left(
\theta_x^2 + \theta_y^2 \right)} ~. ~\label{undurad8bisnew}
\end{eqnarray}
The next step is to calculate the two power fractions corresponding to the $\sigma$ and $\pi$ polarization modes using Eq. (\ref{xpowden}). It is convenient to present the expressions for $W_{\sigma}$ and $W_{\pi}$  as a function of the Fresnel number $N$, defined as

\begin{equation}
N= \frac{\omega_{1o} \sigma^2}{c L_w} ~,\label{frsnintr}
\end{equation}
Since we integrated along $z'$ from $-\infty$ to $0$, we did not need to define explicitly the length of the undulator. However, in order to compare the for $W_{\sigma}$ and $W_{\pi}$  (and more precisely their ratio $W_{\sigma}/W_{\pi}$) with the results obtained above, we still need to present the power as a function of the Fresnel number $N$, defined exactly as before.

Explicit calculations yield

\begin{equation}
\left(
\begin{array}{c}
{W}_{\sigma}\\ {W}_{\pi}
\end{array}\right)= W_o \left(
\begin{array}{c}
\mathrm{A_{JJ}}^2 \rho_{1D}^{-1} G_\sigma(N)\\ \mathrm{B_{JJ}}^2 \rho_{1D} G_\pi(N)
\end{array}\right)~,\label{W02bis}
\end{equation}

where

\begin{eqnarray}
G_\sigma(N) &=& \frac{1}{2 \sqrt{3}} \exp[(1-i\sqrt{3})N] \left\{\pi + \pi \exp\left[2 i \sqrt{3}N\right]\right. \cr && \left. - i \exp\left[2 i \sqrt{3}N\right]\mathrm{Ei}\left(N (-1-i\sqrt{3})\right) +i\mathrm{Ei}\left(i N (i+\sqrt{3})\right)\right\}\cr &&
\label{Gx}
\end{eqnarray}

\begin{eqnarray}
G_\pi(N) &=& \frac{1}{6}\left\{\frac{3}{N} \right.\cr && \left. - \left(-3i +\sqrt{3}\right)\exp[(1+i\sqrt{3})N] \left[\pi - i\mathrm{Ei}\left((-1-i\sqrt{3})N\right) \right] \right.\cr && \left. - \left(3i +\sqrt{3}\right)\exp[(1-i\sqrt{3})N] \left[\pi + i\mathrm{Ei}\left(i(i+\sqrt{3})N\right) \right]\right\}
\label{Gy}
\end{eqnarray}

and

\begin{eqnarray}
W_o = W_b a_{f}^2 \left(\frac{I_o}{\gamma I_A}\right) \left(\frac{K^2}{2+K^2}\right)
\label{Wo}
\end{eqnarray}
with $W_b = m_e c^2 \gamma I_o/e$ is the total power of the electron beam.

The ratio between the fractions radiated in the two modes of polarization is therefore conveniently expressed as a function of three separate factors:

\begin{eqnarray}
\frac{W_\pi}{W_\sigma} = f(K) g(N_w) w(N)
\label{WydivWox2}
\end{eqnarray}
with

\begin{eqnarray}
&& f(K) = \left[\frac{B_{JJ}^2}{A_{JJ}^2}\right] \cr &&
g(\rho_{1D}) = \left[\rho_{1D}^2\right] \cr &&
w(N) = \left[\frac{G_\pi(N)}{G_\sigma(N)}\right]~.
\label{fact2}
\end{eqnarray}

\begin{figure}
\includegraphics[width=0.5\textwidth]{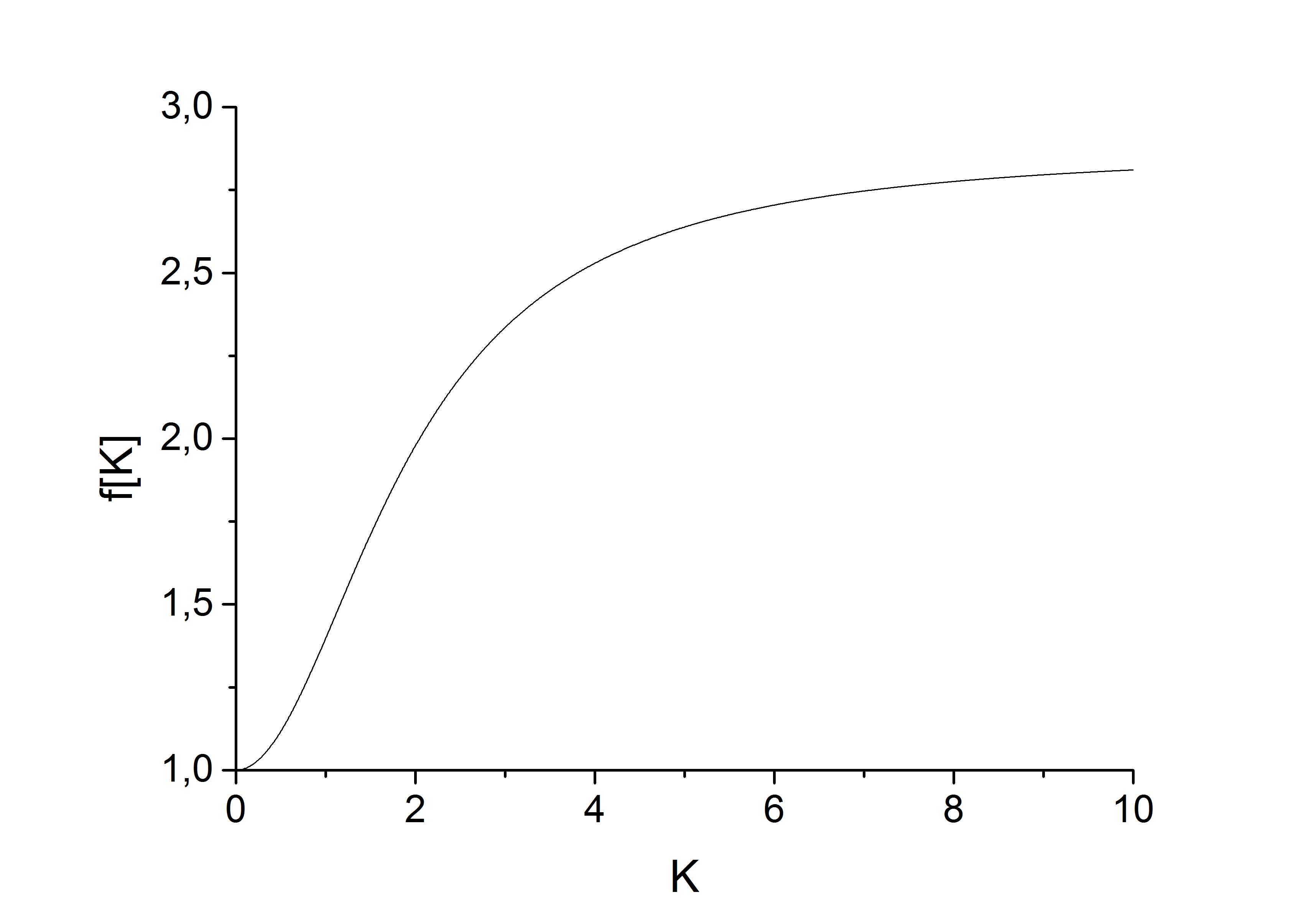}
\includegraphics[width=0.5\textwidth]{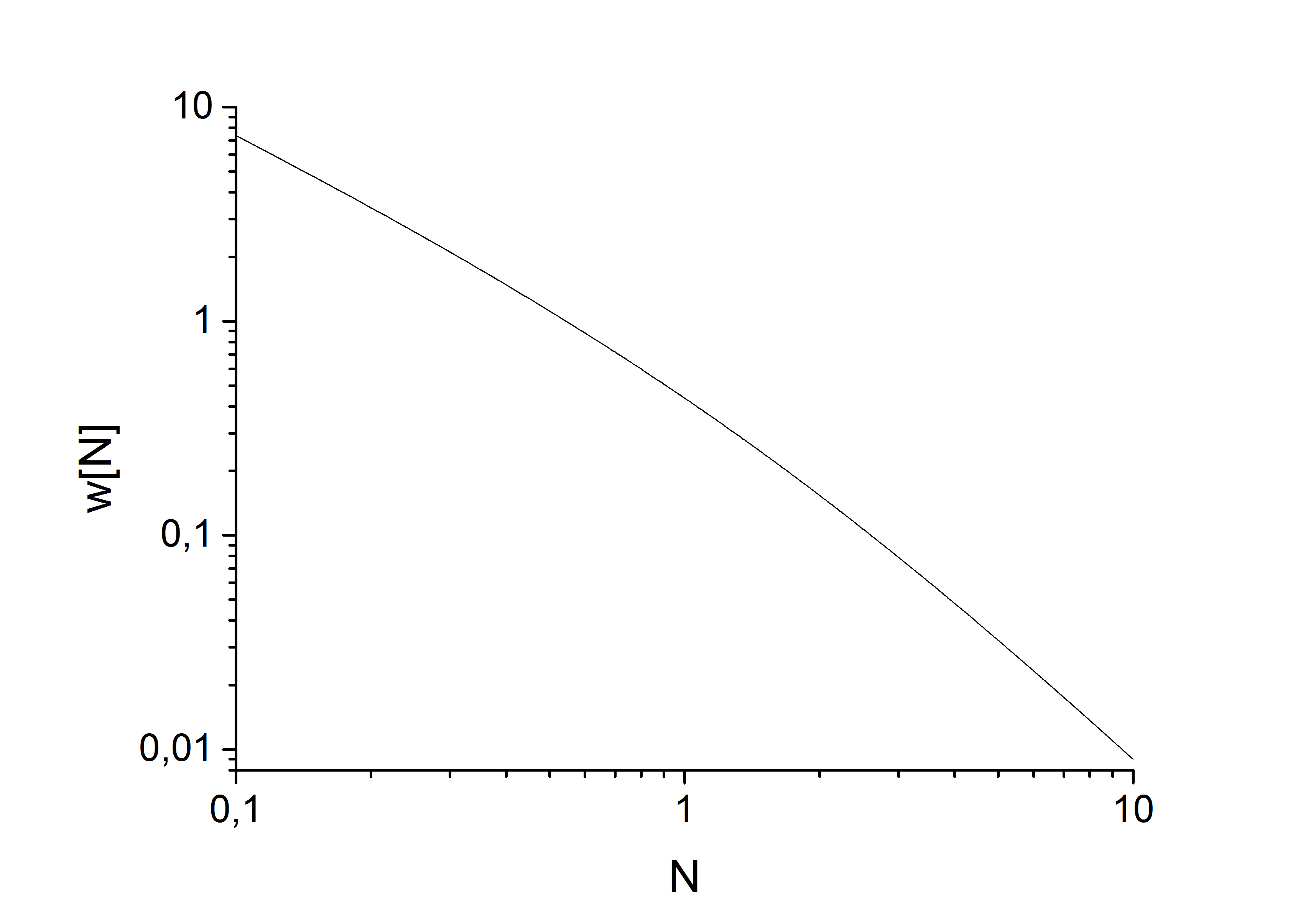}
\caption{Illustration of the behavior of $f(K)$ (left) and $w(N)$ (right).}
\label{f1}
\end{figure}
The first factor, $f(K)$, is only a function of the undulator $K$ parameter and is plotted in Fig. \ref{f1}. As it can be seen from Fig. \ref{f1} it always remains between about unity and $3$, for any value of $K$. The second factor, $g(\rho_{1D})$ scales as the inverse number of undulator periods squared, and is a signature of the fact that the gradient term in the equation for the electric field scales as the inverse number of undulator periods. The third factor, $w(N)$,  is only a function of the diffraction parameter that is, once the wavelength and the undulator length are fixed, a function of the electron beam size only. It is also plotted in Fig. \ref{f1}. It is unity for values of the diffraction parameter around unity, but it quickly decreases for larger values of $N$. Clearly the power fraction radiated in the $\pi$ mode increases drastically with the photon energy, partly due to a larger number of undulator period per field gain-length, but mainly because of a larger diffraction parameter.

For the sake of exemplification we apply the model in the linear regime discussed in this paragraph to one particular case, calculated from start-to-end simulations for the SASE1 and SASE2 line of the European XFEL. We consider a $250$ pC electron beam at a photon energy of about $9$ keV. The electron energy is $17.5$ GeV and $K \simeq 3.6$. The peak current is about $5$ kA, and the rms sizes of the electron beam in the horizontal and vertical directions are about $\sigma_x \simeq 15~\mu$m and $\sigma_y \simeq 18~\mu$m respectively. For our purposes of exemplification\footnote{If the electron beam is not round, one can easily modify the model for the electron beam distribution in this paragraph.} we consider a round beam with $\sigma = 16 \mu$m. The peak current density can then be estimated as $I_0/(2 \pi \sigma^2)$. Finally, the undulator period is $\lambda_w = 40$ mm. From these numbers we obtain the parameter $\rho_\mathrm{1D} \simeq 8 \cdot 10^{-4}$. Plugging these numbers in Eq. (\ref{WydivWox}) and remembering the definition in Eq. (\ref{frsnintr})we obtain $f(K) \simeq 2.5$, $g(\rho_{1D}) \simeq 6.4\cdot 10^{-7}$, $N \simeq 3$ and $w(N) \simeq 0.072$, so that the overall ratio $W_{\pi}/W_{\sigma} \simeq 1.13\cdot 10^{-7}$.

\subsection{Constant density modulation}

In analogy with the previous paragraph, we now proceed to study the case of a constant density modulation along an undulator of fixed length $L_w$, imitating the behavior of an FEL at saturation. We can still set $C(z)=0$. At variance with the previous model we now write

\begin{equation}
\tilde{\rho}({z},\vec{l}) = j_{o}\left(\vec{l}\right) 2 \pi a_f H_{L_g}(z) \delta(\omega-\omega_{1o}) ~. \label{expara}
\end{equation}
Here $a_f = \mathrm{const}$ a constant modulation level, $H_{L_w}(z) = 1$ for $z$ in the range $(-L_w/2,L_w/2)$ and zero otherwise, with $L_w$ the undulator length,  and $j_o$ is defined as in Eq. (\ref{exbot2}). In this case, Eq. (\ref{undurad5}) can be written as

\begin{eqnarray}
\vec{\widetilde{E}}_{\bot1}&=& \frac{2 \pi a_f\omega_{1o}} {c^2 z}
\exp{\left[i \frac{\omega_{1o}}{2c}
{z}({\theta}_x^2+{\theta}_y^2) \right]}
\left[\frac{K}{2\gamma}\mathrm{A_{JJ}}\vec{e}_x +
\frac{2 K \gamma}{2+K^2}\mathrm{B_{JJ}}{\theta}_x {\theta}_y\vec{e}_y\right]\cr&&
\times\int_{-\infty}^{\infty} d {l}_x \int_{-\infty}^{\infty} d
{l}_y \int_{-\infty}^{\infty} d {z}' \exp{\left[-i
\frac{\omega_{1o}}{c} \left({\theta}_x l_x +
\theta_y l_y\right)\right] }  \cr && \times \exp\left[i \frac{\omega_{1o}}{2 c}
\left(\theta_x^2+ \theta_y^2\right) z'\right]
j_{o}\left(\vec{l}\right)  H_{L_w}(z') \delta(\omega-\omega_{1o})
 ~.\label{undurad6}
\end{eqnarray}

The integral in $d \vec{l}$ amounts to the spatial Fourier transform of $j_o\left(\vec{l}\right)$. We proceed similarly as in the previous paragraph and obtain

\begin{eqnarray}
\vec{\mathcal{E}}_{\bot1}&=& -\frac{ I_o a_f\omega_{1o} L_w}
{c^2 z} \exp{\left[i
\frac{\omega_{1o}}{2c}{z}({\theta}_x^2+{\theta}_y^2) \right]} \cr && \times
\left[\frac{K}{2\gamma}\mathrm{A_{JJ}}\vec{e}_x +
\frac{2 K \gamma}{2+K^2}\mathrm{B_{JJ}} {\theta}_x {\theta}_y \vec{e}_y\right] \cr&&
\times\mathrm{sinc} \left[\frac{L_w \omega_{1o}}{4 c
}\left( \theta_x^2 + \theta_y^2 \right)\right] \exp{\left[-\frac{\sigma^2 \omega_{1o}^2 }{2 c^2}\left(\theta_x^2 +  \theta_y^2
\right)\right]} ~. ~\label{undurad8bis}
\end{eqnarray}
The final step, as in the previous paragraph, consists is the calculation of the angle-integrated first harmonic power. As before the power for the \textit{x-} and \textit{y-}polarization components of the first harmonic radiation are given by Eq. (\ref{xpowden}).

Both integrals for the horizontal polarization component and the vertically polarized correction can be calculated analytically by exploiting the cylindrical symmetry of the model. One is then left with

\begin{equation}
\left(
\begin{array}{c}
{W}_{\sigma}\\ {W}_{\pi}
\end{array}\right)= W_o \left(
\begin{array}{c}
\mathrm{A_{JJ}}^2 (4 \pi N_w) F_\sigma(N)\\ \mathrm{B_{JJ}}^2 (4 \pi N_w)^{-1} F_\pi(N)
\end{array}\right)~,\label{W02bis}
\end{equation}

where

\begin{eqnarray}
F_\sigma(N) = \arctan{\left(\frac{1}{{2 N}}\right)}+N\ln{\left(\frac{4 N^2}{4N^2+1} \right)}~,
\label{Fx}
\end{eqnarray}
\begin{eqnarray}
F_\pi(N) = \frac{1}{N(1+4N^2)},
\label{Fy}
\end{eqnarray}

and where parameters $N$ and $W_o$  are given by Eq. (\ref{frsnintr}) and Eq. (\ref{Wo}).

Similarly as before, the ratio between the two fractions radiated into the two modes of polarization is  conveniently expressed as a function of three separate factors:

\begin{eqnarray}
\frac{W_\pi}{W_\sigma} = f(K) g(N_w) h(N)
\label{WydivWox}
\end{eqnarray}
with

\begin{eqnarray}
&& f(K) = \left[\frac{B_{JJ}^2}{A_{JJ}^2}\right] \cr &&
g(N_w) = \left[\frac{1}{(4 \pi N_w)^2}\right] \cr &&
h(N) = \left[\frac{F_\pi(N)}{F_\sigma(N)}\right]~.
\label{facts}
\end{eqnarray}
\begin{figure}
\begin{center}
\includegraphics[width=0.5\textwidth]{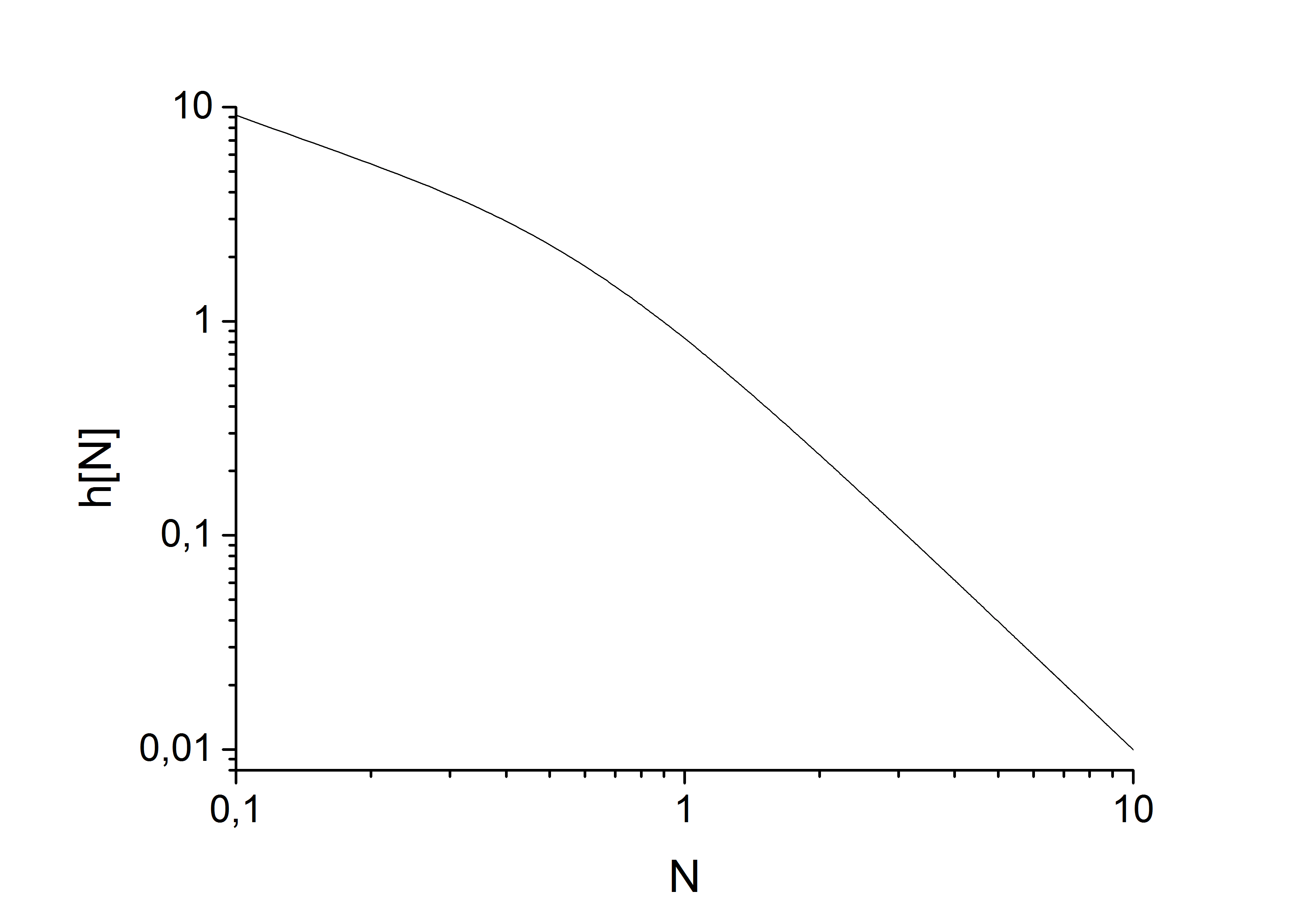}
\caption{Illustration of the behavior of $h(N)$.}
\end{center}
\label{f2}
\end{figure}
The function $f$ has been defined in the previous paragraph. Concerning the second factor $g$, we have an expression which is similar to that in Eq. (\ref{fact2}). The only difference is that here we replaced $\rho_{1D}$ with $(4 \pi N_w)^{-1}$, with $N_w$ the number of undulator periods in the undulator. The number of undulator periods in a field gain length is just $N_w = (4 \pi \rho_\mathrm{1D})^{-1}$, and therefore the second factor in Eq. (\ref{WydivWox}) just amounts to $\rho_\mathrm{1D}^2$ for an undulator length $L_w = L_g$, $L_g$ being, as before, the field gain length. By setting the undulator length equal to the field gain length the two models can be directly compared by studying $w(N)$ as defined in Eq. (\ref{fact2}) and $h(N)$ defined in Eq. (\ref{facts}). We plot $h(N)$ explicitly in Fig. \ref{f2}. As one can see it differs from Fig. \ref{f1}, due to the different model used.

Considering the same example made in the previous paragraph we find again $f(K) \simeq 2.5$, $g(\rho_{1D}) \simeq 6.4\cdot 10^{-7}$, $N \simeq 3$. Plugging the value for $N$ into Eq. (\ref{facts}) we obtain $h(N) \simeq 0.097$, so that the overall ratio $W_{\pi}/W_{\sigma} \simeq 1.5\cdot 10^{-7}$.

\subsection{Polarization characteristics for the second harmonic}

So far we have discussed the values of the two polarization components $\sigma$ and $\pi$. We have shown that, typically, in the case of an XFEL with a horizontal planar undulator, only less that one part in a million of the total power at the first harmonic is polarized in the vertical direction. For some experiments even such small fraction of the $\pi$ mode is of importance.

In addition to the radiation at the first harmonic, however, there are several other background contributions to the vertically polarized power fraction from higher harmonic radiation. It is useful to be able to estimate what is the typical fraction of the $\pi$ mode component at higher harmonics. There is an important distinction to be made between odd harmonics and even harmonics. The $\pi$-mode contribution from
all odd harmonics ($\omega =3 \omega_{1o}$, $\omega = 5 \omega_{1o}$, etc.) can be completely disregarded, while in practice even harmonics ($\omega = 2 \omega_{1o}$, $\omega = 4 \omega_{1o}$, etc.) can still be an important source of vertically polarized radiation. Since the total power contained in each harmonic is proportional to the square of the bunching amplitude,  only the contribution due to the second harmonic is of practical interest.

The contribution of the second harmonic can be calculated using results in \cite{SECO}. We proceed to study the case of a constant density modulation. The ratio between the fraction of total power radiated into $\pi$ mode at $\omega = 2 \omega_{1o}$ and radiated into the $\sigma$-mode at $\omega = \omega_{1o}$ can be expressed as a function of separate factors \cite{SECO}:

\begin{equation}
\frac{W_{2\pi}}{W_{1\sigma}} =  \frac{1}{4 \pi N_w}
\frac{2+K^2}{K^2}\frac{a_{2}^2}{a_{f}^2}
\frac{\mathcal{B}^2}{A_{JJ}^2}
\frac{F_{2\pi}(N)}{F_{\sigma}(N)}~,\label{realcomp}
\end{equation}
where $N$ and $F_{\sigma}$ are given, respectively\footnote{Note that $N$ in \cite{SECO} is defined twice larger with respect to what is reported here. In fact, in \cite{SECO} all results refer to the Fresnel number for the second harmonic.} by Eq. (\ref{frsnintr}) and Eq. (\ref{Fx}), $\mathrm{A_{JJ}}$ is given by
Eq .(\ref{calA}), $a_2$ and $a_f$ are the amplitudes of the beam modulation at $\omega = 2 \omega_{1o}$ and $\omega = \omega_{1o}$ respectively,

\begin{eqnarray}
\mathcal{B} = J_1\left(\frac{K^2}{2+K^2}\right)
\label{mathcalB}
\end{eqnarray}

and

\begin{eqnarray}
F_{2\pi} = \ln{\left(1+\frac{1}{16{N}^2}\right)}
\label{f22222}
\end{eqnarray}

\begin{figure}
\begin{center}
\includegraphics[width=0.5\textwidth]{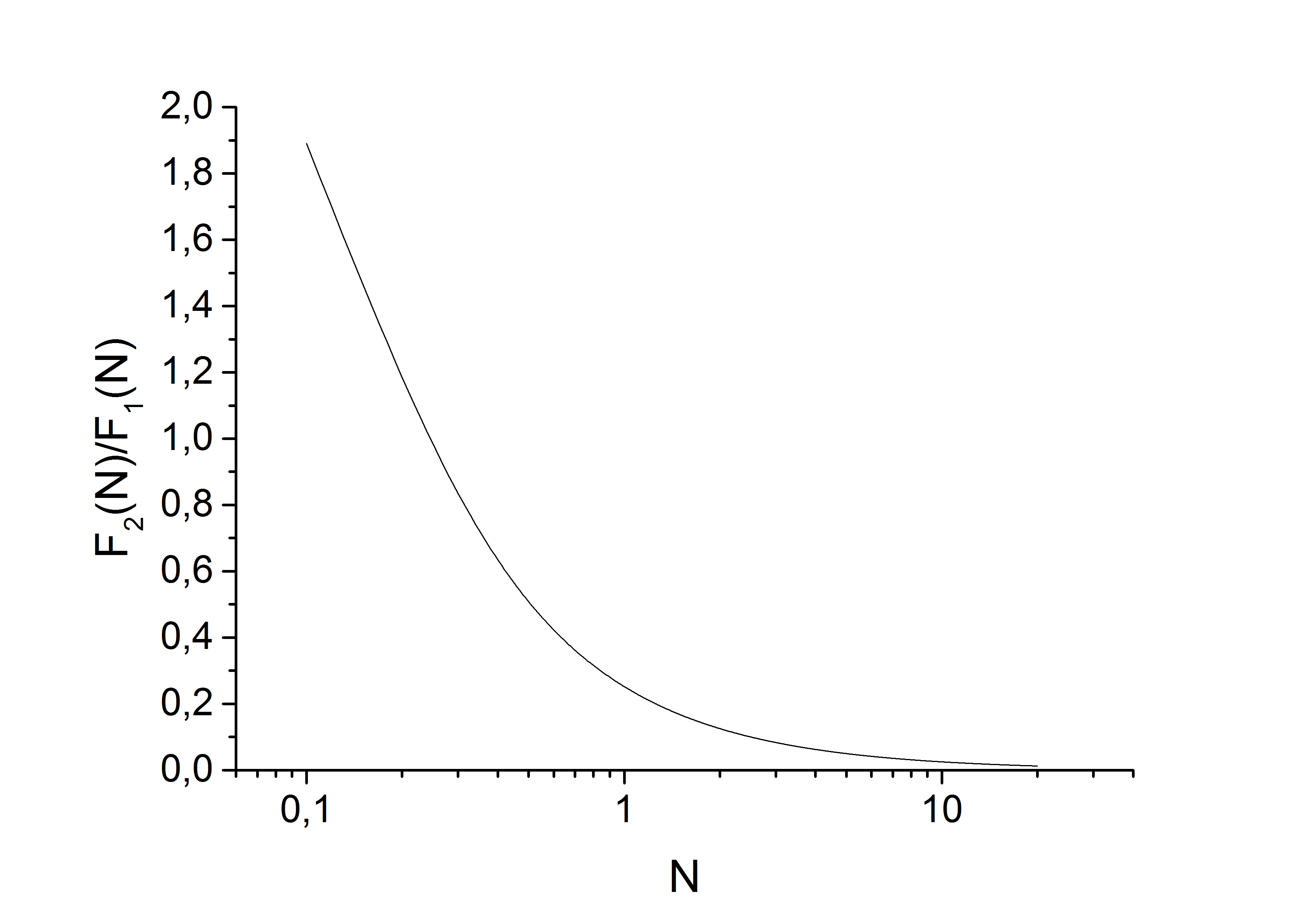}
\caption{Illustration of the behavior of $F_2(N)/F_\sigma(N)$.}
\end{center}
\label{f334}
\end{figure}
We plot $F_2(N)/F_\sigma(N)$ explicitly in Fig. 3.  Considering the same example made above we find

\begin{eqnarray}
\frac{W_{2\pi}}{W_{1\sigma}} \sim 10^{-4} \frac{a_2^2}{a_f^2}
\label{finalratw2w1}
\end{eqnarray}
The contribution from the even harmonics scales therefore as $a_2^2/a_f^2$, and can be completely disregarded when the XFEL operates in linear regime. At saturation, the contribution from the second harmonic can be comparable with the first harmonic in the case when X-ray optics harmonic separation is absent.

\section{Conclusions}

One attractive feature of radiation from X-ray Free-Electron Lasers (XFELs) is its high degree of polarization. This paper shows that for an FEL with a planar undulator with the electron motion on the horizontal plane, the horizontally polarized component of radiation from greatly dominates the photon beam characteristics and only less than one part in a million of the total intensity is polarized in the vertical plane. This feature that can be useful in different experimental situations. When describing physical principles, analytical descriptions are always important: they allow for a proper understanding of the principles, in our case those of FEL physics, and also for testing numerical simulation codes. From this point of view, a SASE XFEL is a rather complicated subject. In fact, in all generality,  its radiation can be represented as a non-stationary random process whose analytical description is complicated by the fact that the electron bunch combines both the features of the input signal and of the active medium with time-dependent parameters. Approximations are therefore needed. In particular, it is important to find a model allowing for an analytical description without loss of essential information about the features of the FEL process. A model satisfying these conditions is that of a stepped profile electron bunch, together with the application of self-seeding scheme for narrowing the radiation bandwidth. In the framework of this model it becomes possible to describe the polarization properties of the radiation from an XFEL in a fully analytical way. Using Maxwell's equations one can write an explicit expression for calculating the electric field with given charge and current sources. Up to now, in all FEL codes the contribution of the charge term is assumed to be negligibly small. However, in our case of interest, the charge term is the only responsible for the vertically polarized radiation component. Our analytical results, in particular those for the high-gain linear regime, are therefore expected to serve as a primary standard for testing future FEL codes including the charge term as electromagnetic source.

\section{\label{sec:ackn} Acknowledgements}

The authors wish to thank Tom Cowan for stimulating discussions, Serguei Molodtsov and Thomas Tschentscher for their support and their interest in this work.

\newpage

\bibliographystyle{elsarticle-num}
\bibliography{bib_1_GG}

%\begin{thebibliography}{99}
%\bibitem{SECO} G. Geloni, E. Saldin, E. Schneidmiller and M.
%Yurkov, [[[EXACT SOLUTION]]]
%\bibitem{FELP} E. Saldin, E. Schneidmiller and M. Yurkov, The
%Physics of Free Eelectron Lasers, Springer, 2000
%\bibitem{KIM1} Z. Huang and K. Kim, Phys. Rev. E, 62, 5 (2000)
%\bibitem{SCHM} M. Schmitt and C. Elliot, Phys. Rev. A, 34, 6
%(1986)
%\bibitem{KIM2} Z. Huang and K. Kim, Nucl. Instr. and Meth. in
%Phys. Res. A 475, 112 (2001)
%\bibitem{FREU} H. Freund, S. Biedron and S. Milton., Nucl. Instr. and Meth. in
%Phys. Res. A 445, 53 (2000)
%\bibitem{TREM} A. Tremaine et al., Phy. Rev. Lett. 88, 204801
%(2002)
%\bibitem{BIED} S. Biedron et al. Nucl. Instr. Meth. A 483, 94
%(2002)
%\bibitem{RHUA} Z. Huang and S. Reiche, Proceedings of the FEL 2004
%Conference, Trieste, Italy
%\bibitem{WIE2} H. Wiedemann, Synchrotron Radiation,
%Springer-Verlag, Germany (2003)
%\bibitem{UNDU} Undulators, Wigglers and their applications, Edited
%by H. Onuki and P. Ellaume, Taylor $\&$ Francis (2003)
%\bibitem{OURS} G. Geloni, E. Saldin, E. Schneidmiller and M.
%Yurkov, Paraxial Green's functions in Synchrotron Radiation
%theory, DESY 05-032, ISSN 0418-9833 (2005)
%\bibitem{ALFE} D. Alferov, Y.A. Bashmakov et al.
% Sov. phys. - Tech. Phys. 18, 1336 (1974)
%\bibitem{METH} E. Saldin et al. Nucl. Instr. Meth. A 539, 499 (2005)
%\end{thebibliography}

\end{document}